\documentclass[12pt]{article}
\def\bea{\begin{eqnarray}}
\def\ve{\vert}
\def\eea{\end{eqnarray}}

\def\nnb{\nonumber}
\def\la{\langle}
\def\ra{\rangle}
\def\ba{\begin{array}}
\def\ea{\end{array}}
\def\tr{\mbox{Tr}}
\def\ssp{{\Sigma^{*+}}}
\def\sso{{\Sigma^{*0}}}
\def\ssm{{\Sigma^{*-}}}
\def\xis0{{\Xi^{*0}}}
\def\xism{{\Xi^{*-}}}
\def\ss{\la \bar s s \ra}
\def\uu{\la \bar u u \ra}

\def\gGgG{\la g^2 G^2 \ra}
\def\om{{\Omega^-}}
\def\structure{g_{\mu \nu}\! \not\!p_1\!\!\not\!\epsilon\not\!p_2}
\def\fo{f_0(\frac{s_0}{M^2})}
\def\ffi{f_1(\frac{s_0}{M^2})}

\begin{document}
\title{\bf Magnetic Moments of Decuplet Baryons in Light Cone QCD}
\author{T.M. Aliev, A. {\"O}zpineci} 
\begin{titlepage}
\maketitle
\begin{abstract}

We calculate the magnetic moments of decuplet  baryons containing strange quarks within the framework of
light cone QCD sum rules taking into account the $SU(3)$ flavor symmetry breaking effects. It is
obtained that magnetic moments of the neutral $\sso$ and $\xis0$ baryons are mainly determined by the
$SU(3)$ breaking terms. A comparison of our results on the magnetic moments of the decuplet baryons with
the predictions of other approaches is presented.
\end{abstract} 
\end{titlepage}

\section{Introduction}
For the determination of the fundamental parameters of hadrons from experiments, some information
about physics at large distances is required.  The large distance physics can not be calculated
directly from fundamental QCD Lagrangian because at large distance perturbation theory can not be
applied.  For this reason a reliable non-perturbative approach is needed.  Among  non-perturbative
approaches, QCD sum rules \cite{shifman} occupied a special place  in studying the properties of
ground state hadrons. This method is applied to various problems in hadron physics  and extended in many
works (see for example Refs. \cite{reinders, ioffe1, shifman22} and references therein). The magnetic
moments of hadrons are one of their characteristic parameters in low energy physics. Calculation of the
nucleon magnetic moments in the framework of QCD sum rules method using external fields technique, first
suggested in \cite{ioffe2}, was carried out in \cite{ioffe3, balitsky1}.  They were later refined and
extended to the entire baryon octet in \cite{chiu, pasupathy}.

In \cite{belyaev2,lee1}, magnetic moments of the decuplet baryons are calculated  within the framework
of QCD sum rules using external field method.  Note that in \cite{belyaev2}, from the decuplet baryons,
only the magnetic moments of $\Delta^{++}$ and $\Omega^-$ were calculated.  At present, the magnetic
moments of $\Delta^{++}$ \cite{bosshard}, $\Delta^{0}$ \cite{heller} and $\Omega^-$
\cite{wallace} are known from experiments.  The experimental information provides new incentives for
theoretical  scrutiny of these physical quantities.

Recently, we have calculated the magnetic moments of the $\Delta$ baryons \cite{aliev} within the
framework of an alternative approach to the traditional sum rules, i.e. the light cone QCD sum rules
(LCQSR). In this work, the magnetic moments of other members of the decuplet which contain at least one 
$s$-quark, namely the $\Sigma^{*\pm,0}$, $\Xi^{*0,-}$ and $\Omega^-$, are calculated within the same 
approach. The novel feature of the present work is that we take into account the $SU(3)$ flavor symmetry
breaking effects.

 A few words about the LCQSR method are in order. The LCQSR is based on the operator product
expansion on the light cone, which is an expansion over the
twists of the operators rather than dimensions as in the traditional QCD sum rules.  The main
contribution comes from the lower twist operator.  The matrix elements of the nonlocal operators between
the vacuum and hadronic state defines the hadronic wave functions. (More about this method and its
applications can be found in \cite{braun2, braun22} and references therein).  Note that magnetic moments
of the nucleon using LCQSR approach was studied in \cite{braun3}.

The paper is organized as follows.  In Sect. II, the light cone QCD sum rules for the magnetic moments
of the decuplet baryons are derived.  In Sect. III, we carry out numerical calculations. Comparison of
the  predictions of this approach on the magnetic moments of the decuplet baryons with the results of
other methods, and the experimental results is also presented in this section.

\section{Sum Rules for the Magnetic Moments of Decuplet Baryons}
A sum rule for the magnetic moment can be constructed by equating two different representations of the
corresponding correlator, written in terms of hadrons and quark-gluons. We  begin our calculations by
considering the
following correlator:
\bea
\Pi_{\mu \nu} = i \int dx e^{i p x} \la 0 \ve {\cal T} \eta^B_\mu(x) \bar \eta^B_\nu(0) \ve 0
\ra_F
\label{correlation}\, ,
\eea
where ${\cal T}$ is the time ordering operator, $F$ means electromagnetic field and the
$\eta^B_\mu$'s are the interpolating currents of the corresponding baryon,  B, carrying the same quantum
numbers.  This correlator can be calculated on one side phenomenologically, in terms of the
hadron parameters, and on the other side by the operator product expansion (OPE) in the deep Eucledian
region, $p^2 \rightarrow - \infty$, using QCD degrees of freedom.
By equating both expressions, we construct the corresponding sum rules.

Saturating the correlator, Eq. (\ref{correlation}), by ground state baryons we get:
\bea
\Pi_{\mu \nu}(p_1^2,p_2^2) = \frac{\la 0 \ve \eta^B_{\mu} \ve
B_1(p_1) \ra}{p_1^2 - M_1^2} 
\la B_1 (p_1) \ve  B_2(p_2) \ra_F 
\frac{\la B_2 (p_2) \ve \eta^B_\nu \ve 0 \ra}{p_2^2 - M_2^2},
\label{insert}
\eea 
where $p_2 = p_1 + q$, $q$  is the photon momentum and $M_i$ is the mass of the baryon $B_i$.

The matrix elements of the interpolating currents between the ground state and
the state containing a single baryon, $B$, with momentum $p$ and having spin $s$ is defined as:
\bea
\la 0 \ve \eta_{\mu} \ve B(p,s) \ra = \lambda_B u_\mu (p,s) \, , \label{3}
\eea
where $\lambda_B$ is the residue, and $u_\mu$ is the Rarita-Schwinger spin-vector (For a discussion of
the properties of the Rarita-Schwinger spin-vector see e.g. \cite{taka}).
In order to write down the phenomenological part of the sum rules from Eq. (\ref{insert}) it follows
that one also needs an expression for the matrix element $\la B(p_1) \ve B (p_2) \ra_F$, i.e. the
electromagnetic vertex of spin $3/2$ baryons.   In the general case, 
this vertex can be written as:
\bea
\la B(p_1) \ve B(p_2) \ra_F = \epsilon_\rho \bar u_\mu(p_1) {\cal O}^{\mu \rho
\nu}(p_1,p_2) u_\nu (p_2) \, , \label{4}
\eea
where $\epsilon_\rho$ is the polarization vector of the photon and the Lorentz tensor ${\cal O}^{\mu
\rho \nu}$ is given by:
\bea
{\cal O}^{\mu \rho \nu}(p_1,p_2) &=& - g^{\mu \nu} \left[ \gamma_\rho (f_1+f_2) +
\frac{(p_1+p_2)_\rho}{2 M_B}
f_2 +q_\rho f_3 \right]  - \nnb \\
&-& \frac{q_\mu q_\nu}{(2 M_B)^2} \left[ \gamma_\rho (G_1+G_2) + \frac{(p_1 + p_2)_\rho}{2 M_B} G_2 +
q_\rho
G_3 \right] \label{5}
\eea
where the form factors $f_i$ and $G_i$ are functions of $q^2=(p_1 - p_2)^2$.  In
our problem, the values of the formfactors only at one point, $q^2=0$, are needed.

In calculations,  summation over spins of the Rarita-Schwinger spin vector is performed,
\bea
\sum_s u_\sigma(p,s) \bar u_\tau(p,s) = - \frac{(\not\! p + M_B)}{2 M_B} \left\{ 
g_{\sigma \tau} - \frac{1}{3} \gamma_\sigma \gamma_\tau - \frac{2 p_\sigma p_\tau}{3 M_B^2} +
\frac{p_\sigma \gamma_\tau - p_\tau \gamma_\sigma}{3 M_B} \right\} \label{6}
\eea
Using Eqs. (\ref{insert}-\ref{6}), one can see that 
the correlator contains many structures, not all of them
independent.  To remove the dependencies, an ordering of the gamma matrices should be chosen.   
For this purpose the ordering $\gamma_\mu\!\!\not\!p_1\!\!\not\!\epsilon\!\!\not\!p_2 \gamma_\nu$ 
is chosen.  With this ordering, the correlation function becomes:
\bea
\Pi_{\mu \nu} &=& \lambda_B^2 \frac{1}{(p_1^2 - M_B^2)(p_2^2 - M_B^2)} \left[ g_{\mu \nu} 
\not\!p_1\!\not\!\epsilon\!\not\!p_2
\frac{g_M}{3} + \right. \nnb \\
&+&  \mbox{\it other structures with $\gamma_{\mu}$ at the beginning and  $\gamma_{\nu}$ at the
end \huge $]$}
\eea 
where $g_M$ is the magnetic form factor, $g_M/3 = f_1 +f_2$.  The value of $g_M$ at $q^2 =0$
gives the magnetic moment of the baryon in units of its natural magneton, $e \hbar/2 m_B c$.
Hence, among the many structures in the correlator, for determination of the magnetic moments, only the
structure $g_{\mu \nu}\not\!p_1\!\!\not\!\epsilon\!\!\not\!p_2$ is needed. The appearance of the factor
$3$ can be understood from the fact that
in the nonrelativistic limit, the maximum energy of the baryon in the presence of a uniform magnetic
field with magnitude $H$  is $3(f_1 + f_2) H \equiv g_M \, H$ \cite{belyaev1}.  
Another advantage of choosing the $g_{\mu \nu}\not\!p_1\!\!\not\!\epsilon\!\!\not\!p_2$ structure is
that spin $1/2$ baryons do not contribute to this structure.
Indeed, their overlap is given by: 
\bea
\la 0 \ve \eta_\mu \ve J=1/2 \ra = (A p_\mu + B \gamma_\mu) u(p)
\eea
where $(\not\!p - m) u(p) = 0$ and $(A m + 4 B) = 0$ \cite{belyaev1,leinweber1}, and we can not
construct the structure $g_{\mu \nu}\not\!p_1\!\!\not\!\epsilon\!\!\not\!p_2$.

For calculating the correlator (\ref{correlation}) from the QCD side, first of all,
suitable interpolating currents should be chosen.  For the baryons under study, they can be chosen
as (see for example \cite{lee1}): 
\bea
\eta_\mu^{\Sigma^{*+}} &=& \frac{1}{\sqrt{3}} \epsilon^{abc} [2 (u^{aT} C \gamma_\mu s^b) u^c +
(u^{aT} C \gamma_\mu u^b) s^c] \, , \nnb \\
\eta_\mu^{\Sigma^{*0}} &=&\sqrt{\frac{2}{3}} \epsilon^{abc} [ (u^{aT} C \gamma_\mu d^b) s^c +
(d^{aT} C \mu_\alpha s^b) u^c + (s^{aT} C \gamma_\mu u^b) d^c] \, , \nnb \\
\eta_\mu^{\Sigma^{*-}} &=& \frac{1}{\sqrt{3}} \epsilon^{abc} [2 (d^{aT} C \gamma_\mu s^b) d^c +
(d^{aT} C \gamma_\mu d^b) s^c] \, , \nnb \\
\eta_\mu^{\Xi^{*0}} &=& \frac{1}{\sqrt{3}} \epsilon^{abc} [2 (s^{aT} C \gamma_\mu u^b) s^c +
(s^{aT} C \gamma_\mu s^b) u^c] \, , \nnb \\
\eta_\mu^{\Xi^{*-}} &=& \frac{1}{\sqrt{3}} \epsilon^{abc} [2 (s^{aT} C \gamma_\mu d^b) s^c +
(s^{aT} C \gamma_\mu s^b) d^c] \, , \nnb \\
\eta_\mu^{\Omega^-} &=& \epsilon^{abc}  (s^{aT} C \gamma_\mu s^b) s^c 
\eea
where $C$ is the charge conjugation operator, $a, \, b,\, c$ are color indices.
It should be noted that these baryon currents are not unique, one can choose  an
infinite number of currents with the same quantum numbers \cite{ioffe4,chung}.

After some calculations, for the theoretical parts of the correlator, we get:
\bea
\Pi^\ssp_{\mu \nu} &=& {\Pi'}^\ssp_{\mu \nu}
- \frac{1}{6} \epsilon^{abc} \epsilon^{def} \int d^4 x
e^{ipx}\la \gamma(q) \ve \bar u^d A_i u^a \nnb \\
&&\left\{
2 A_i \gamma_\nu {S'_s}^{be} \gamma_\mu S_u^{cf} +
2 A_i \gamma_\nu {S'_u}^{cf} \gamma_\mu S_s^{be} +
\right. \nnb \\
&& \left. +
2 S_s^{be} \gamma_\nu A'_i \gamma_\mu S_u^{cf} +
2 A_i \tr (\gamma_\nu {S'_u}^{cf} \gamma_\mu S_s^{be}) +
\right. \nnb \\
&& \left. +
 S_s^{be} \tr (\gamma_\nu A'_i \gamma_\mu S_u^{cf}) +
\right. \nnb \\
&& \left. +
2 S_u^{cf} \gamma_\nu {S'_s}^{be} \gamma_\mu A_i +
2 S_u^{cf} \gamma_\nu A'_i \gamma_\mu S_s^{be} +
\right. \nnb \\
&& \left. +
2 S_s^{be} \gamma_\nu {S'_u}^{cf} \gamma_\mu A_i +
2 S_u^{cf} \tr (\gamma_\nu A'_i \gamma_\mu S_s^{be}) +
\right. \nnb \\
&& \left. +
S_s^{be} \tr(\gamma_\nu {S'_u}^{cf} \gamma_\mu A_i)
\right\} + \bar s^e A_i s^b \nnb \\
&&\left\{
2 S_u^{ad} \gamma_\nu A'_i \gamma_\mu S_u^{cf} +
2 S_u^{ad} \gamma_\nu {S'_u}^{cf} \gamma_\mu A_i +   
\right. \nnb \\
&& \left. +
2 A_i  \gamma_\nu {S'_u}^{ad} \gamma_\mu S_u^{cf} +
2 S_u^{ad} \tr (\gamma_\nu {S'_u}^{cf} \gamma_\mu A_i) +
\right. \nnb \\
&& \left. +
A_i \tr (\gamma_\nu {S'_u}^{ad} \gamma_\mu S_u^{ad} )
\right\} \ve 0 \ra \label{sigmasp} \\
\Pi_{\mu \nu}^{\om} =&& \Pi_{\mu \nu}'^{\om}   
+\frac{1}{2} \epsilon^{abc} \epsilon^{def} \int d^4 x e^{ipx}\la \gamma(q) \ve \bar s^f A_i s^a \nnb \\
&&\left\{
2 S_s^{cd} \gamma_\nu S_s'^{be} \gamma_\mu A_i +
2 S_s^{cd} \gamma_\nu A'_i \gamma_\mu S_s^{be} +
\right. \nnb \\
&& \left.
+ 2 A_i \gamma_\nu S_s'^{cd} \gamma_\mu S_s^{be} +
S_s^{cd} \tr (\gamma_\nu {S'_s}^{be} \gamma_\mu A_i) +
\right. \nnb \\
&&\left.
+S_s^{cd} \tr (\gamma_\nu  A'_i \gamma_\mu S_s^{be}) +
A_i \tr (\gamma_\nu {S'_s}^{cd} \gamma_\mu S_s^{be})
\right\} \ve 0 \ra \label{omegam} 
\eea
where $A_i = 1, \, \gamma_\alpha, \, \sigma_{\alpha \beta}/\sqrt{2}, \, i \gamma_\alpha \gamma_5,
\, \gamma_5$, a sum over $A_i$ implied, $S' \equiv CS^TC$, $A'_i = CA_i^TC$, with $T$ denoting the
transpose of the matrix, and $S_q$ is the full light quark propagator with both perturbative and
non-perturbative contributions.  We calculate the theoretical part of the sum rules in linear
order in the strange quark mass, $m_s$. The calculations show that, the terms quadratic in the
strange quark mass give  smaller contributions than the terms linear in $m_s$ (about $8\%$). For the
propagator of quarks, we will use the following expression: 
\bea
S_q &=& \la 0 \ve {\cal T} \bar q(x) q(0) \ve 0 \ra \nnb \\
&=& \frac{i \not\!x}{2 \pi^2 x^4} - \frac{m_q}{4 \pi^2 x^2} - 
\frac{\la \bar q q \ra}{12}\left( 1- \frac{i m_q}{4} \not\!x \right) 
 - \frac{x^2}{192}m_0^2 \la \bar q q \ra \left(1- \frac{i m_q}{6} \not\!x
\right) - \nnb \\ 
&-& i g_s \int_0^1 dv \left[ \frac{\not\!x}{16 \pi^2 x^2} G_{\mu \nu}(v x)  \sigma_{\mu \nu} - v x_\mu
G_{\mu \nu}(v x) \gamma_\nu \frac{i}{4 \pi^2 x^2} - \right. \nnb \\
&-&\left. \frac{i m_q}{32 \pi^2} G_{\mu \nu} \sigma_{\mu \nu} \left( \ln \frac{-x^2 \Lambda^2}{4} + 2
\gamma_E \right) \right] \label{12}
\eea  
where $\Lambda$ is an energy cutoff separating perturbative and non-perturbative regimes.

In Eqs. (\ref{sigmasp})-(\ref{omegam}), the first terms, $\Pi_{\mu \nu}'^B$, describe diagrams in
which the
photon interact with the quarks perturbatively.  Their explicit expressions can be obtained from
the remaining terms by substituting all occurances of 
\bea
\bar q^a(x) A_i q^b {A_i}_{\alpha \beta} \rightarrow 2 \left( \int d^4 y  F_{\mu \nu} y_\nu S_q^{pert}
(x-y) \gamma_\mu S_q^{pert}(y)\right)^{ba}_{\alpha \beta} \label{pert}
\eea
where the Fock-Schwinger gauge, $x_\mu A_\mu(x)=0$ is used, and $S_q^{pert}$ is the perturbative part 
of the quark propagator, i.e. the first two terms in Eq. (\ref{12}). Here, $F_{\mu \nu}$ is the
electromagnetic field strength tensor.

For customary, here we presented theoretical results only for the correlators of $\ssp$ and $\om$ (see
Eqs. (\ref{sigmasp}) and (\ref{omegam})). The corresponding expressions for the theoretical
parts of the correlators for the $\ssm$, $\sso$, $\xis0$ and $\xism$ baryons can be obtained from
Eq. (\ref{sigmasp}) as follows: For $\ssm$, substitute $d$ quarks instead of $u$ quarks; for $\xis0$
exchange $u$ and $s$ quarks; and for $\xism$, substitute $s$ quarks instead of $u$ quarks, and $d$
quarks instead of $s$ quarks. The theoretical part of the 
correlator for the $\sso$ baryon is half the sum of the theoretical parts of the correlators for
the $\ssp$ and $\ssm$ baryons in exact $SU(2)$ flavor symmetry limit. 

For calculating the QCD part of the sum rules, one needs to know the matrix elements $\la \gamma (q) \ve
\bar q A_i q \ve 0 \ra$.  Upto twist-4,  matrix elements contributing to the selected
$g_{\mu \nu} \not\!p_1\!\not\!\epsilon\!\not\!p_2$ structure are expressed in terms of the photon wave
functions as \cite{balitsky, braun, ali}: 
\bea
\la \gamma(q) \ve \bar q \gamma_\alpha \gamma_5 q \ve 0 \ra &=& \frac{f}{4} e_q  \epsilon_{\alpha \beta
\rho \sigma} \epsilon^\beta q^\rho x^\sigma \int_0^1 du e^{i u q x} \psi(u) \nnb \\
\la \gamma(q) \ve \bar q \sigma_{\alpha \beta} q \ve 0 \ra &=& i e_q  \la \bar q q \ra \int_0^1 du e^{i
u q x} \nnb \\
&\times& \{ (\epsilon_\alpha q_\beta - \epsilon_\beta q_\alpha) [\chi \phi(u) + x^2 [ g_1(u) - g_2(u) ]]
\nnb \\
&+& \left[ q x (\epsilon_\alpha x_\beta - \epsilon_\beta x_\alpha) + \epsilon x (x_\alpha q_\beta -
x_\beta
q_\alpha ) \right] g_2(u) \} \label{13}
\eea
where $\chi$ is the magnetic susceptibility of the quark condensate and $e_q$ is the quark charge.
The functions $\phi(u)$ and $\psi (u)$ are the leading twist-2 photon wave functions, while $g_1(u)$ and
$g_2 (u)$ are the twist-4 functions.

Using Eqs. (\ref{12}) and (\ref{13}), after some algebra, and performing Fourier transformation,  the
result for the structure  $g_{\mu \nu} \not\!p_1 \not\!\epsilon\not\!p_2$  can be obtained. As stated
earlier, in order to construct the sum rules, we must equate the
phenomenological and theoretical expressions for the correlator. Performing the Borel transformation on
the variables $p^2$ and $(p+q)^2$ in order to suppress the contributions of the higher resonances and the
continuum, the following sum rules for the magnetic moment of the baryons are obtained:
\bea
g_M^\ssp &=& \frac{e^{\frac{M_{\Sigma^*}^2}{M^2}}}{\lambda_{\Sigma^*}^2} \left\{
\frac{f \psi(u_0)}{12 \pi^2} \left[ \frac{\la g^2 G^2 \ra}{48} - 
M^4 f_1(\frac{s_0}{M^2}) \right] (e_s + 2 e_u) + \right. \nnb \\
&+& \left. 
\frac{8}{3} \la \bar u u \ra (g_1(u_0)-g_2(u_0)) \left[ \la \bar s s \ra (e_s+e_u) + \la \bar u u \ra
e_u \right] +
\right. \nnb \\
&+& \left.
\frac{\chi \phi(u_0) \la \bar u u \ra}{6} \left[ m_0^2 - 4 M^2 f_0(\frac{s_0}{M^2}) \right] ( \ss (e_s +
e_u) + \uu e_u) +
\right. \nnb \\
&+& \left. 
\frac{2}{3} \la \bar u u \ra (e_s \la \bar u u \ra + 2 e_u \la \bar s s \ra)+
\frac{\gGgG M^2}{768 \pi^4} f_0(\frac{s_0}{M^2}) (e_s + 2 e_u) + \right. \nnb \\
&+& \left. 
\frac{3 M^6}{64 \pi^2} f_2(\frac{s_0}{M^2}) (e_s + 2 e_u) +
\frac{m_s M^2}{4 \pi^2} f_0(\frac{s_0}{M^2})(e_u \ss - e_s \uu)- 
\right. \nnb \\
&-& \left.
\frac{m_s \uu}{8 \pi^2} \left(\gamma_E - \ln\frac{\Lambda^2}{M^2}\right) \left[ m_0^2 e_s + \frac{
e_u}{9} \gGgG \phi(u_0) \chi \right]+
\right. \nnb \\
&+& \left. 
\frac{m_s \uu M^2}{\pi^2} f_0(\frac{s_0}{M^2}) \left[ e_s \gamma_E - 2 e_u (g_1(u_0)-g_2(u_0)) -e_u
\right]+
\right. \nnb \\
&+& \left. 
\frac{e_u m_s \uu}{4 \pi^2} \left(m_0^2 - \frac{2}{9} (g_1(u_0)-g_2(u_0)) \frac{\gGgG}{M^2} +
\right. \right. \nnb
\\
&+& \left. \left.
\frac{8}{3}  \pi^2 f \psi(u_0) + \chi \phi(u_0) M^4 f_1(\frac{s_0}{M^2}) \right)
\right\} \, , \label{srssp} \\
g_M^\xis0 &=& \frac{e^{\frac{M_{\Xi^*}^2}{M^2}}}{\lambda_{\Xi^*}^2} \left\{
\frac{f \psi(u_0)}{12 \pi^2} \left[ \frac{\la g^2 G^2 \ra}{48} - 
M^4 f_1(\frac{s_0}{M^2}) \right] (e_u + 2 e_s) + \right. \nnb \\
&+& \left. 
\frac{8}{3} \la \bar s s \ra (g_1(u_0)-g_2(u_0)) \left[ \la \bar u u \ra (e_u+e_s) + \la \bar s s \ra
e_s \right]
+ \right. \nnb \\
&+& \left. 
\frac{\chi \phi(u_0) \la \bar s s \ra}{6} \left[ m_0^2 - 4 M^2 f_0(\frac{s_0}{M^2}) \right] (\uu (e_u +
e_s) + \ss e_s ) +
\right. \nnb \\
&+& \left. 
\frac{2}{3} \la \bar s s \ra (e_u \la \bar s s \ra + 2 e_s \la \bar u u \ra)+
\frac{\gGgG M^2}{768 \pi^4} f_0(\frac{s_0}{M^2}) (e_u + 2 e_s) + \right. \nnb \\
&+& \left. 
\frac{3 M^6}{64 \pi^2} f_2(\frac{s_0}{M^2}) (e_u + 2 e_s) +
\right. \nnb \\
&-& \left.
\frac{m_s}{\pi^2}(g_1(u_0)-g_2(u_0)) (\ss e_s + \uu e_u) \left(2 M^2 \fo + \frac{\gGgG}{18 M^2}
\right) -
\right. \nnb \\
&-& \left.
\frac{m_s \chi \phi(u_0)}{72 \pi^2} \gGgG (\ss e_s + \uu e_u) \left( \gamma_E - \ln
\frac{\Lambda^2}{M^2}
\right) +
\right. \nnb \\
&-& \left.
\frac{m_0^2 m_s e_s}{8 \pi^2} (\ss + \uu) \left( \gamma_E - \ln \frac{\Lambda^2}{M^2}
\right) +
\right. \nnb \\
&+& \left.
m_s \left( \frac{2}{3} f \psi(u_0) + \frac{m_0^2}{4 \pi^2} \right) (\uu e_s + \ss e_u) 
\right. \nnb \\
&+& \left.
\frac{m_s}{4 \pi^2} M^2 \fo \left[ 4 \gamma_E e_s (\ss +\uu) - (5 \uu e_s + 3 \ss e_u) \right]
\right. \nnb \\
&+& \left.
\frac{m_s \chi \phi(u_0)}{4 \pi^2} M^4 \ffi (\ss e_s + \uu e_u)
\right\} \label{srxis0} \, ,\\
g_M^\om &=&  \frac{e_s}{\lambda_{\Omega}^2} e^{\frac{M_{\Omega}^2}{M^2}}\left\{ \frac{f
\psi(u_0)}{4 \pi^2} \left[\frac{\la g^2 G^2 \ra}{48} -  M^4 f_1(\frac{s_0}{M^2}) \right] +
\right. \nnb \\
&+& \left. 8 \la \bar s s \ra^2 [g_1(u_0) - g_2(u_0)] + \right. \nnb \\
&+& \left. \frac{\chi \phi(u_0) \la \bar s s \ra^2}{2}
\left[m_0^2 -4 M^2 f_0(\frac{s_0}{M^2}) \right] \right. \nnb \\
&+& \left.  2 \la \bar s s \ra^2 +\frac{\la g^2 G^2 \ra M^2}{256
\pi^4}f_0(\frac{s_0}{M^2}) + \frac{9 M^6}{64 \pi^4}f_2(\frac{s_0}{M^2}) +
2 f \psi(u_0) m_s \ss - 
\right. \nnb \\
&-& \left.
\frac{m_s \ss}{6 \pi^2} \gGgG \left[ \frac{g_1(u_0)-g_2(u_0)}{M^2} + \chi
\phi(u_0) \left(
\gamma_E - \ln \frac{\Lambda^2}{M^2} \right) \right] - \right. \nnb \\
&-& \left.
\frac{6}{\pi^2} m_s \ss (g_1(u_0) - g_2(u_0)) M^2 \fo + 
\right. \nnb \\
&+& \left.
\frac{3 m_0^2}{8 \pi^2} m_s \ss \left(2 -
\gamma_E + \ln
\frac{\Lambda^2}{M^2} \right) - \right. \nnb \\
&-& \left.
\frac{3(1-\gamma_E)}{\pi^2} m_s \ss M^2 \fo + \frac{3 \chi \phi(u_0)}{4 \pi^2} m_s \ss M^4 \ffi
\right\} \label{srom} \, .
\eea
As is stated earlier, the sum rules for $\Sigma^{*\pm}$, and $\xism$ can be obtained from
Eq. (\ref{srssp}) and Eq. (\ref{srxis0}), respectively as follows: To obtain the sum rules for $\ssm$
and $\sso$ from Eq. (\ref{srssp}), replace $e_u$ by $e_d$ and $(e_u + e_d)/2$ respectively.  To
obtain the sum rules for $\xism$, replace $e_u$ by $e_d$ in Eq. (\ref{srxis0}).

In Eqs. (\ref{srssp})-(\ref{srom}), the functions
\bea
f_n(x) = 1 - e^{-x} \sum_{k=0}^n \frac{x^k}{k!}
\eea
are used to subtract the contributions of the continuum and $s_0$ is the continuum
threshold,
\bea
u_0 &=& \frac{M_2^2}{M_1^2 + M_2^2} \nnb \\
\frac{1}{M^2} &=& \frac{1}{M_1^2} + \frac{1}{M_2^2} \nnb
\eea
As we are working with just a single baryon, the Borel parameters $M_1^2$ and $M_2^2$ should be taken to
be equal, i.e. $M_1^2 = M_2^2$, from which it follows that $u_0 = 1/2$.

\section{Numerical Analysis}

From the sum rules, one sees that,  besides several constants, one needs expressions for the photon wave
functions in  order to calculate the numerical value of the magnetic moment of
the decuplet baryons. It was shown in \cite{balitsky, braun} 
that they do not deviate much from the asymptotic form, hence, we shall use the following photon wave
functions \cite{braun,ali}:
\bea
\phi(u) &=& 6 u \bar u \nnb \\
\psi(u) &=& 1 \nnb \\
g_1(u) &=& -\frac{1}{8} \bar u (3 - u) \nnb \\
g_2(u) &=& - \frac{1}{4} \bar u ^2 \nnb
\eea
where $\bar u = 1-u $.
The values of the other constants that are used in the calculation are: $f=0.028 \, GeV^2$, $\chi = - 
4.4 \, GeV^{-2}$ \cite{belyaev4} (in \cite{balitsky2}, $\chi$ is estimated to be $\chi = -3.3 \,
GeV^{-2}$), $\la g^2 G^2 \ra = 0.474 \, GeV^4$, $\la \bar u u \ra = \ss / 0.8= - (0.243)^3 \, GeV^3$,
$m_0^2 = (0.8 \pm 0.2)\, GeV^2$ \cite{belyaev3}, $\lambda_{\Sigma^*} = 0.043 \, GeV^3$, 
$\lambda_{\Xi^*} = 0.053 \, GeV^3$, $\lambda_{\Omega} = 0.068 \, GeV^3$ \cite{lee5}. For the
energy cut-off, $\Lambda$, we will take $\Lambda=0.5 \, GeV$.

Having fixed the input parameters, our next task is to find a region of Borel parameter, $M^2$, where
dependence of the magnetic moments on $M^2$ and the continuum threshold $s_0$ is rather weak and at the
same time higher states and continuum contributions remain under control.
We demand that these contributions  are less then $35\%$.  Under this requirement, the working region 
for the Borel parameter, $M^2$, is found to be $1.1\, GeV^2 \le M^2 \le 1.4 \, GeV^2$ for $\Sigma^*$
baryons and $1.1 \, GeV^2 \le M^2 \le 1.7 \, GeV^2$ for $\Xi^*$ and $\Omega^-$ baryons. In the case of
$\Xi^*$ and $\Omega^-$ baryons, the working  region of the Borel parameter is wider due to the
relatively large masses of these baryons. 

In Figs. 1-6, we present the dependence of the magnetic moment of each baryon on the Borel parameter,
$M^2$ for three values of the continuum threshold and for the cases $m_s=0$ and $m_s=0.15 \, GeV$.  The
magnetic moments  depend weekly on the value of the continuum threshold, they change at most $6\%$
by a variation of $s_0$ and are also very weakly dependent on $M^2$. From these figures we can deduce
the following conclusions. When we take into account mass of strange quarks, the results for the
magnetic moments of charged decuplet baryons change about $25\%$, but for the neutral decuplet
baryons, the situation changes drastically, i.e. the results increase by more than a factor of
four.  This fact can be explained in the following way.  In exact $SU(3)$ limit, magnetic moments of
$\sso$ and $\xis0$ are proportional to $(e_u+e_d+e_s)$ and $(e_u+2e_s)$, respectively. For example, the
$\xis0$ case is evident from Eq. (\ref{srxis0}) if in  this equation we put $m_s \rightarrow 0$ and $\uu
= \ss$.  In other words, magnetic moments of $\sso$ and $\xis0$ are exactly zero in $SU(3)$ symmetry
limit. (In Figs. 2 and 4, they are slightly different from zero .  This is due to the fact that in the
calculations we take $\ss \ne \la \bar q q \ra$ ($q=u,d$)).  Hence, the main contribution to the
magnetic moments of $\sso$ and $\xis0$ come from $SU(3)$ breaking terms (the mass of $s$-quark,
$s$-quark condensate, etc.).  For this reason, for the magnetic moments of the neutral decuplet baryons
$SU(3)$ breaking effects play an essential role. Note that, all the graphs are  plotted for   $\chi =
-4.4 \, GeV^2$  and $m_0^2=0.8 \, GeV^2$. Our final results on the magnetic moments of the decuplet
baryons at $m_s=0.15 \, GeV$ is presented in Table 1.  For completeness, in this table, we also depicted
our previous predictions on the magnetic moments of $\Delta$ baryons and also the predictions of other
methods.  The quoted errors in Table 1, are  due to the uncertainties in $m_0^2$, $s_0$, variation of
the Borel parameter $M^2$ and the neglected $m_s^2$ terms. One final remark is that our predictions on
the magnetic moment of $\xis0$ differ from the QCD sum rule results not just in magnitude, but also,
more essentially, by sign.
\newpage
\section*{Appendix A}
In this appendix, derivation of the rules for Fourier and Borel transformation which we have used in our
calculations will be presented.

In coordinate representation, the structures that contribute to the structure $\structure$ are
$x_\mu x_\nu \not\!x\not\!\epsilon\not\!q$ and $g_{\mu \nu} \not\!x\not\!\epsilon\not\!q$.
Let us start with the following expressions:
\bea
\int d^4 x e^{i P x} \frac{x_\mu x_\nu  x_\alpha}{(-x^2)^n} \label{four1}
\eea
and
\bea
\int d^4 x e^{i P x} \frac{ x_\alpha}{(-x^2)^n} \label{four2}
\eea
for arbitrary $n$ (there are also terms proportional to $\ln(-x^2)$, these terms will be discussed
later). Note that we are interested only in the part of the Fourier transforms that are proportional to 
$g_{\mu \nu}$. In Eqs. (\ref{four1}) and (\ref{four2}), $P^2=(p+u q)^2 = p_1^2 \bar u + p_2^2 u$ where
$\bar
u = 1-u$. The derivation will be demonstrated for Eq. (\ref{four1}), as generalization is quite trivial.
One can replace every occurance of $x_\beta$ by $-i \frac{\partial}{\partial P_\beta}$.
\bea
\int d^4 x e^{i P x} \frac{x_\mu x_\nu  x_\alpha}{(-x^2)^n} &=& 
\left(-i \frac{\partial}{\partial P_\alpha} \right)
\left(-i \frac{\partial}{\partial P_\mu} \right)
\left(-i \frac{\partial}{\partial P_\nu} \right)
\frac{(-i)}{\Gamma(n)} \times \nnb \\ 
&\times& \int d^4 x \int_0^\infty dt e^{-i P x} t^{n-1} e^{-t x^2} \label{first}
\eea
where we have switched to the Euclidean space in the integral and used the identity
\bea
\frac{1}{y^n} = \frac{1}{\Gamma(n)} \int_0^\infty t^{n-1} e^{-t y}
\eea 
In Eq. (\ref{first}) one should be careful in taking the derivatives as the derivatives are with respect
to the Minkowskian four vector $P$ but the integrand is expressed in terms of the Euclidean vector $P$.
The four dimensional integral is now a trivial Gaussian integration. After performing the integration
over Euclidean space time, and taking the derivatives, the coefficient of $g_{\mu \nu} P_\alpha$ is
found to be
\bea
\int d^4 x e^{i P x} \frac{x_\mu x_\nu  x_\alpha}{(-x^2)^n} \rightarrow
\frac{\pi^2}{4 \Gamma(n)} \int_0^\infty dt t^{n-5} e^{-\frac{P^2}{4 t}}
\eea
Using the Borel transformation of the exponential
\bea
B_{p_1^2} B_{p_2^2} e^{-\frac{P^2}{4 t}} = 
\delta \left(\frac{1}{M_1^2} - \frac{\bar u}{4 t} \right)
\delta \left(\frac{1}{M_2^2} - \frac{u}{4 t} \right)
\eea
and carrying out the $t$ integration, one obtains
\bea
\int d^4 x e^{i P x} \frac{x_\mu x_\nu  x_\alpha}{(-x^2)^n} \rightarrow
\frac{\pi^2}{\Gamma(n)} \left(\frac{M^2}{4}\right)^{n-3} M^2 \delta (u-u_0)
\eea
where 
\bea
M^2 &=& \frac{M_1^2 M_2^2}{M_1^2+M_2^2} \nnb \\
u_0 &=& \frac{M_1^2}{M_1^2 + M_2^2} \nnb
\eea
Similarly
\bea
\int d^4 x e^{i P x} \frac{x_\alpha}{(-x^2)^n} &\rightarrow&
- \frac{2 \pi^2}{\Gamma(n)} \left(\frac{M^2}{4}\right)^{n-2} M^2 \delta (u-u_0) \\
\int d^4 x e^{i P x} \frac{\ln (-x^2) x_\alpha}{(-x^2)^n } &\rightarrow&
-\frac{2 \pi^2}{\Gamma(n)} \left(\frac{M^2}{4}\right)^{n-2} M^2 \times \nnb \\
&\times&\left\{ \ln \left(\frac{M^2}{4} \right) - \frac{d}{dn} \ln \Gamma(n) \right\}\delta (u-u_0) \\
\int d^4 x e^{i P x} \frac{\ln(-x^2) x_\mu x_\nu  x_\alpha}{(-x^2)^n} &\rightarrow&
\frac{ \pi^2}{\Gamma(n)} \left(\frac{M^2}{4}\right)^{n-3} M^2 \times \nnb \\
&\times&\left\{ \ln \left(\frac{M^2}{4} \right) - \frac{d}{dn} \ln \Gamma(n) \right\}\delta (u-u_0)
\eea

The corresponding transformation rules for terms containing $\ln(-x^2)$ have been obtained by making use
of the identity
\bea
\ln(-x^2) = \left. - \frac{\partial}{\partial \epsilon} \frac{1}{(-x^2)^\epsilon} \right|_{\epsilon=0} 
\eea
\newpage

\newpage
\section*{Figure Captions}
\begin{itemize}
\item[Fig. 1.] The dependence of the magnetic moment of $\ssp$ on the Borel parameter, $M^2$, (in units
of nuclear magneton) for three different values of the continuum threshold, $s_0$, and for the cases
$m_s=0$ and $m_s=0.15 \, GeV$.
\item[Fig. 2.] The same as Fig. 1, but for $\sso$.
\item[Fig. 3.] The same as Fig. 1, but for $\ssm$.
\item[Fig. 4.] The same as Fig. 1, but for $\xis0$.
\item[Fig. 5.] The same as Fig. 1, but for $\xism$.
\item[Fig. 6.] The same as Fig. 1, but for $\om$.
\end{itemize}

\newpage
$\left. \right.$
\begin{figure}
    \includegraphics{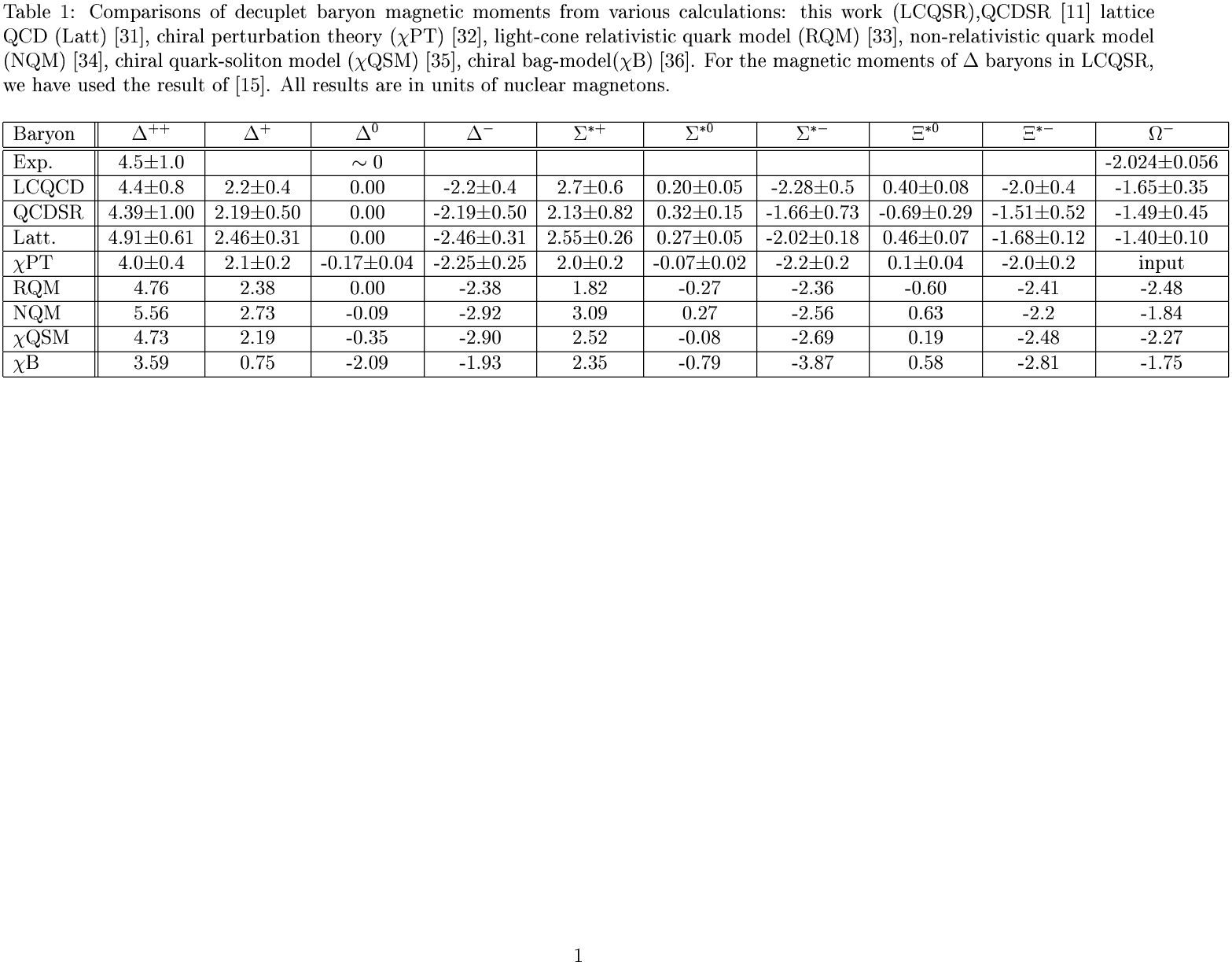}
$\left. \right.$
\end{figure}

\newpage
\begin{figure}
\vskip 8.0 cm
    \includegraphics{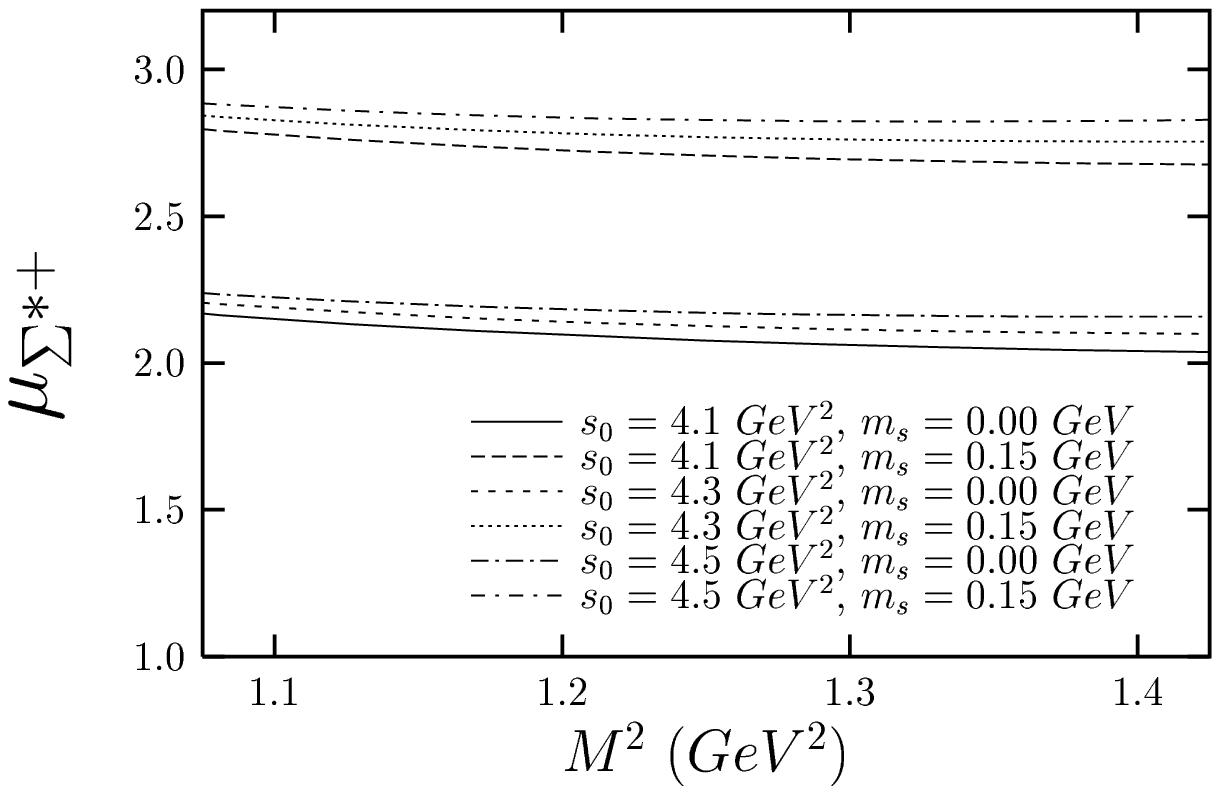}
\vskip -2.5cm
\caption{}
$\left. \right.$
\vspace{2.5cm}
\end{figure}
\begin{figure}
$\left. \right.$
\vskip 4.5 cm
    \includegraphics{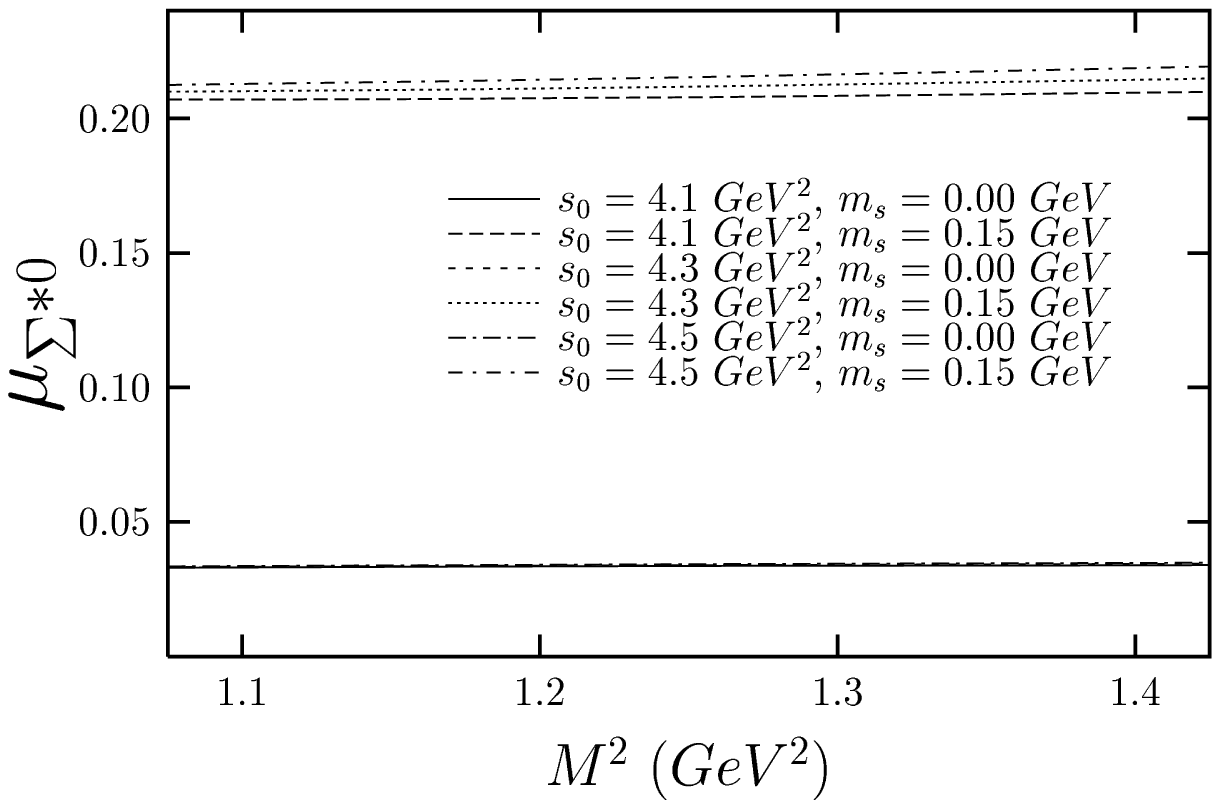}
\vskip -2.5cm
\caption{}
\end{figure}
\newpage
\begin{figure}
\vskip 7.5 cm
    \includegraphics{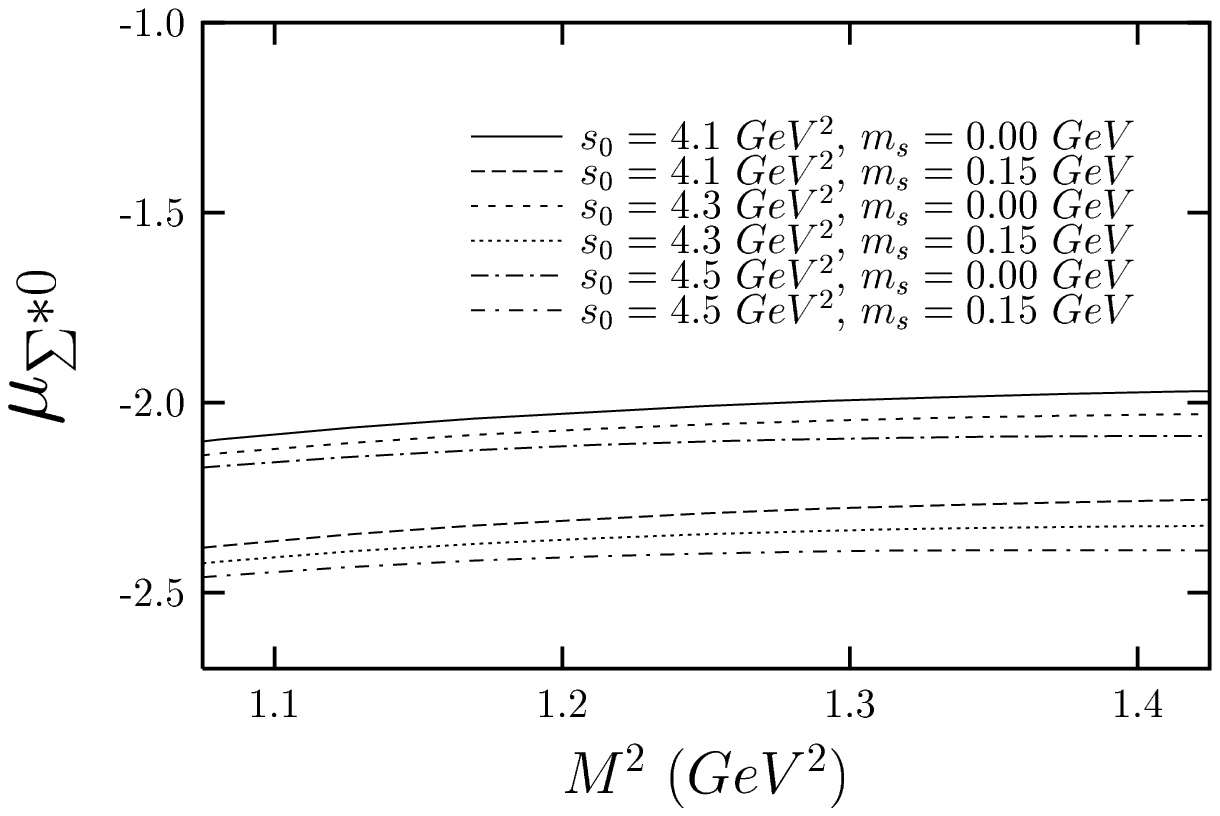}
\vskip -2.1cm
\caption{}
\vspace{2cm}
\end{figure}
\begin{figure}
\vskip 7.5 cm
    \includegraphics{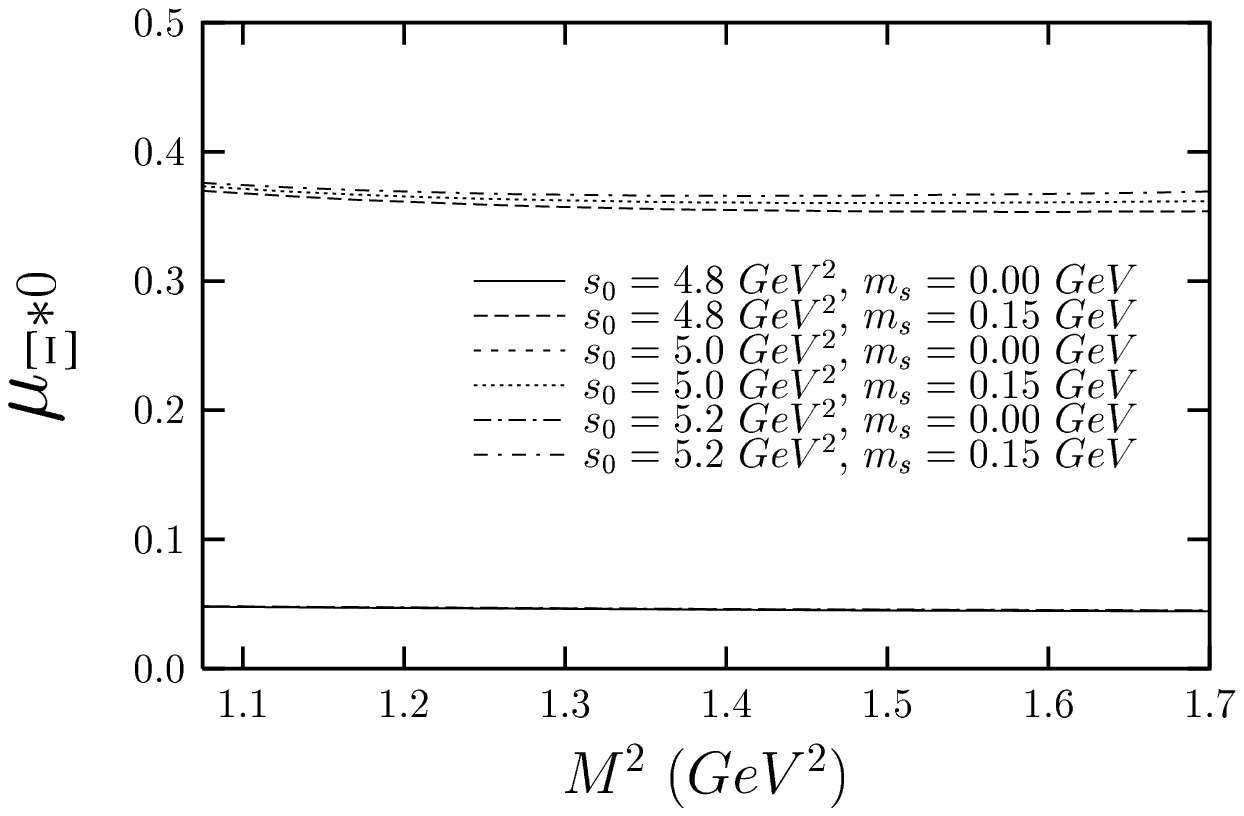}
\vskip -2.1cm
\caption{}
\end{figure}
\newpage
\begin{figure}
\vskip 7.5 cm
    \includegraphics{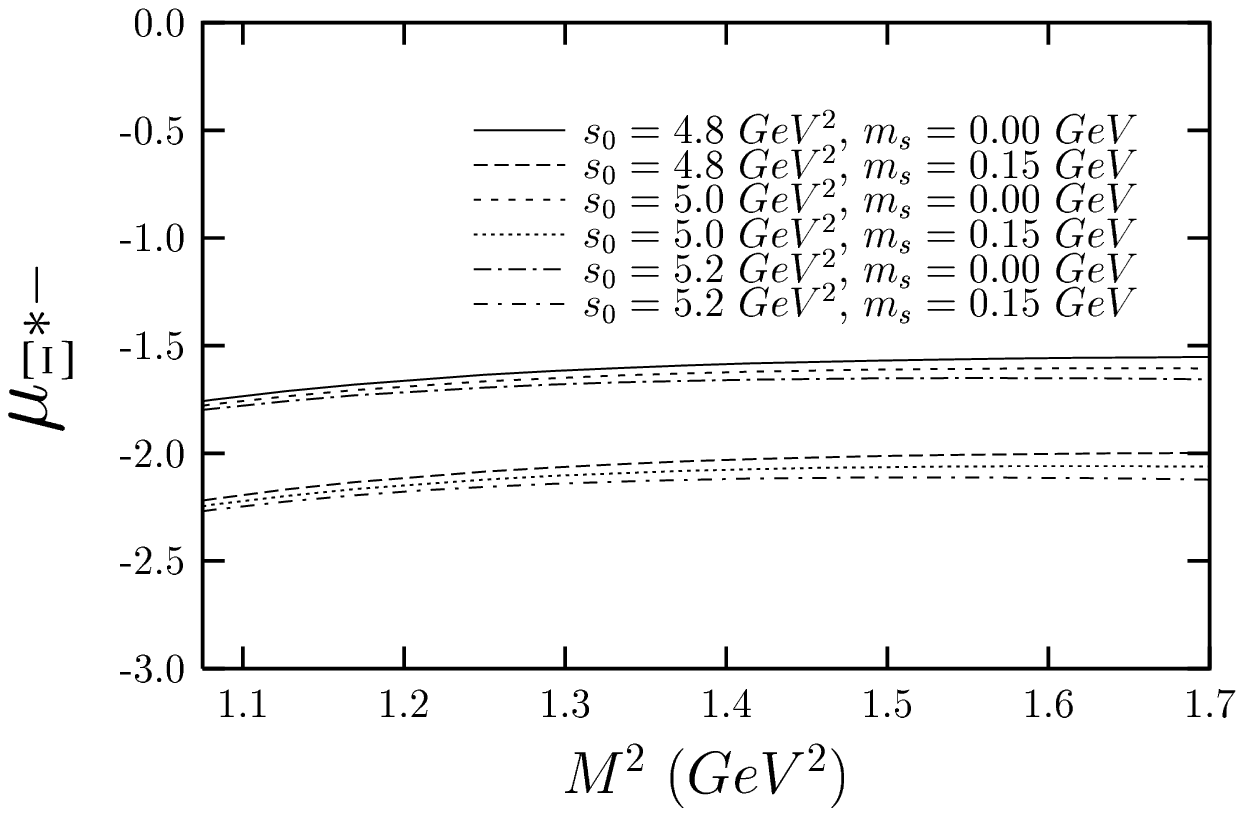}
\vskip -2.1cm
\caption{}
\vspace{2cm}
\end{figure}
\begin{figure}
\vskip 7.5 cm
    \includegraphics{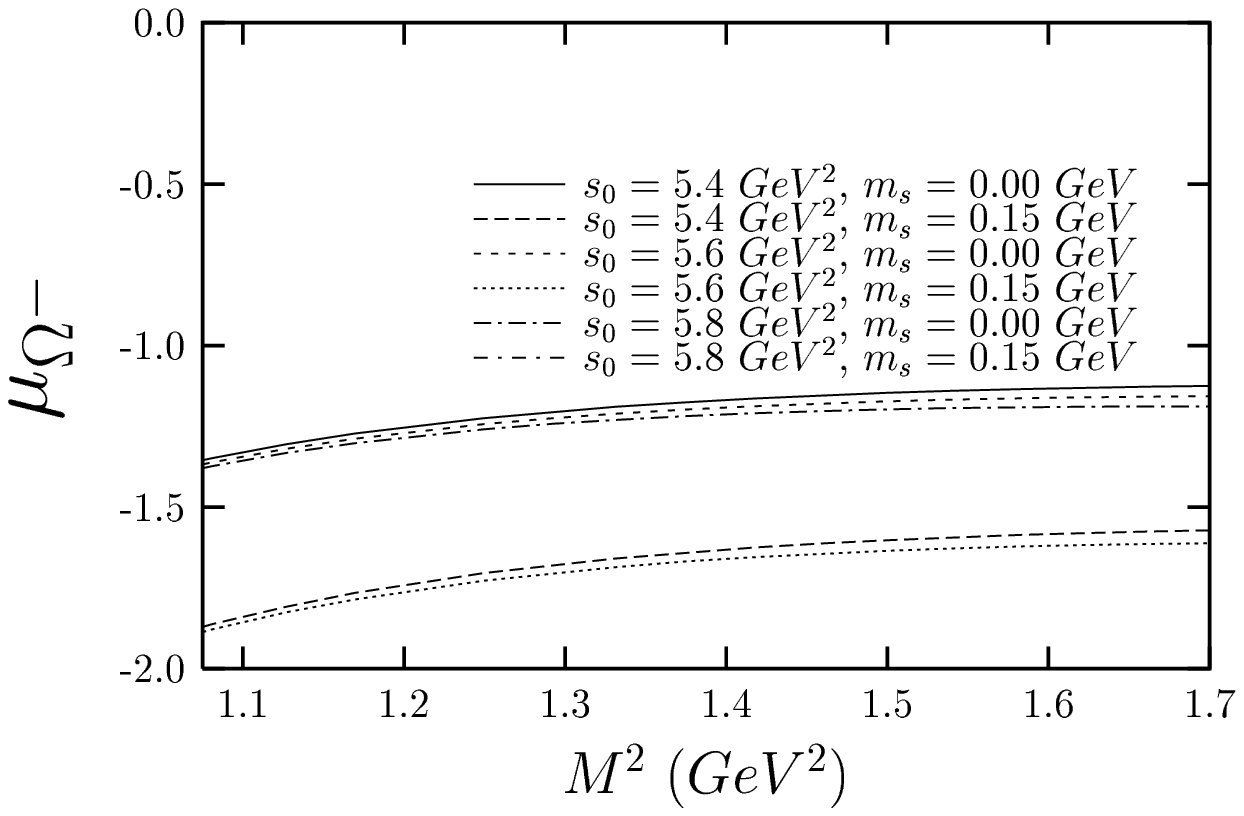}
\vskip -2.1cm
\caption{}
\end{figure}


\begin{thebibliography}{99}

\bibitem{shifman} M. A. Shifman, A. I. Vainstein, and V. I. Zakharov, Nucl Phys. {\bf B147}, 385 (1979)

\bibitem{reinders} L. I. Reinders, H. R. Rubinstein, and S. Yazaki, Phys. Rep. {\bf 127C}, 1 (1985)

\bibitem{ioffe1} "Vacuum Structure and QCD Sum Rules" M. A. Shifman (ed.) North-Holland 1992

\bibitem{shifman22} M. A. Shifman, Prog. Theor. Phys. Supp. 131 (1998) 1

\bibitem{ioffe2} B. L. Ioffe, and A. V. Smilga, JETP Lett  {\bf 37} (1983) 250 

\bibitem{ioffe3} B. L. Ioffe, and A. V. Smilga, Nucl. Phys. {\bf B232} (1989) 109

\bibitem{balitsky1} I. I. Balitsky, and A. V. Yung, Phys. Lett {\bf 1290} (1983) 328

\bibitem{chiu} C. B. Chiu, J. Pasupathy, S. L. Wilson, and C. B. Chiu, Phys. Rev. {\bf D33}, 1961 (1986)

\bibitem{pasupathy} J. Pasupathy, J. P. Singh, S. L. Wilson, and C. B. Chiu, Phys. Rev. {\bf D36}, 1457
(1987)

\bibitem{belyaev2} V. M. Belyaev, prep. ITEP-118 (1984) 

\bibitem{lee1} F. X. Lee, Phys.Rev. {\bf D57}, 1801 (1998) 


\bibitem{bosshard} A. Bosshard {\it et al.}, Phys. Rev. {\bf D44}, 1962 (1991)

\bibitem{heller} L. Heller, S. Kumano, J.C. Martinez, and E.J. Moniz, Phys. Rev. {\bf C35}, 718 (1987)

\bibitem{wallace} N. B. Wallace {\it et al.}, Phys. Rev. Lett. {\bf 74} (1995) 3732

\bibitem{aliev} T. M. Aliev, A. Ozpineci and M. Savci, prep. hep-ph/0002228

\bibitem{braun2} V. M. Braun, Proc. Rostock 1997, "Progress on Heavy Quark Physics" p.105 (1997)

\bibitem{braun22} V. M. Braun, prep. hep-ph/9911206

\bibitem{braun3} V. M. Braun and I. E. Filyanov, Z. Phys {\bf C44} (1989) 157


\bibitem{taka} Y. Takahashi, {\it An Introduction to Field Quantization} (Pergamon Press, London, 1969)

\bibitem{belyaev1} V. M. Belyaev, prep. hep-ph/9301257

\bibitem{leinweber1} D. R. Leinweber, T. Draper, and R. M. Woloshyn, Phys. Rev {\bf D46}, 3067 (1992)

\bibitem{ioffe4} B. L. Ioffe, Z. Phys. {\bf C18} (1983) 67

\bibitem{chung} V. Chung, H. G. Dosch, M. Kremer, and D. Scholl, Nucl. Phys. {\bf B197}, 55 (1982)

\bibitem{balitsky} I. I. Balitsky, V. M. Braun, and A. V. Kolesnichenko, Nucl. Phys. {\bf B312}, 509
(1989)

\bibitem{braun} V. M. Braun and I. E. Filyanov, Z. Phys. {\bf C48}, 239 (1990)

\bibitem{ali} A. Ali and V. M. Braun, Phys. Lett. {\bf B359}, 223 (1995)

\bibitem{belyaev4} V. M. Belyaev and Ya. I. Kogan, Yad. Fiz. {\bf 40} (1984) 1035

\bibitem{balitsky2} I. I. Balitsky and A. V. Kolesnichenko, Yad. Fiz. {\bf 41} (1985) 282 

\bibitem{belyaev3} V. M. Belyaev and B. L. Ioffe, JETP {\bf 56} 493,1982

\bibitem{lee5} F. X. Lee, Phys.Rev. {\bf C57} (1998) 322

\bibitem{Derek92} D.B. Leinweber, T. Draper, and R.M. Woloshyn,                 
Phys. Rev. {\bf D46}, 3067 (1992).

\bibitem{Butler94} M.N. Butler, M.J. Savage,  and R.P. Springer,
Phys. Rev. {\bf D49}, 3459 (1994).

\bibitem{Schlumpf93} F. Schlumpf, Phys. Rev. {\bf D48}, 4478 (1993).

\bibitem{PDG92} Particle Data Group, Phys. Rev. {\bf D45}, 1 (1992).

\bibitem{Kim97} H.C. Kim, M. Praszalowicz, and K. Goeke, Phys.Rev. {\bf D57} (1998) 2859

\bibitem{hong} S. T. Hong, D. P. Min prep. hep-ph/9909004

\end{thebibliography}
\end{document}